\documentclass[aps,pre,preprint,showpacs,floatfix]{revtex4}
\usepackage{graphicx,bm,epsfig}
\begin{document}
\title{Geometric properties of two-dimensional O($n$) loop configurations}
\author{Chengxiang Ding$^{1}$, Xiaofeng Qian$^{2}$, 
Youjin Deng$^{3}$, Wenan Guo$^{1}$
and Henk W.J.~Bl\"ote$^{2,4}$ 
}
\affiliation{$^{1}$ Physics Department, Beijing Normal University, 
Beijing 100875, P.R.China\\
$^{2}$ Instituut Lorentz, Universiteit Leiden,
Postbus 9506, 2300 RA Leiden, The Netherlands \\
$^{3}$ New York University, USA\\
$^{4}$Faculty of Applied Sciences, Delft University of
Technology,  P.O. Box 5046, 2600 GA Delft, The Netherlands \\
}
\date{\today}
\begin{abstract}
We study the fractal geometry of O($n$) loop configurations
in two dimensions by means of scaling and a Monte Carlo method, and
compare the results with predictions based on the Coulomb gas technique.
The Monte Carlo algorithm is applicable to models with
noninteger $n$ and uses local updates. Although these updates typically
lead to nonlocal modifications of loop connectivities, the number of 
operations required per update is only of order one. The Monte Carlo
algorithm is applied to the O($n$) model for several values of $n$, 
including noninteger ones. We thus determine scaling exponents that
describe the fractal nature of O($n$) loops at criticality. The results
of the numerical analysis agree with the theoretical predictions.
\end{abstract}
\maketitle

\section{Introduction}
The O($n$) model consists of $n$-component spins  
$\vec{s_{i}}=( s^{1}_{i},s^{2}_{i}, \ldots s^{n}_{i})$ on a lattice, with
isotropic, i.e., O($n$) invariant couplings. The common form of the reduced
Hamiltonian of the O($n$) spin model is
\begin{equation}\label{spinhamilton}
H=-\frac{J}{k_B T} \sum_{\langle i, j\rangle } \vec{s_{i}} \cdot \vec{s_{j}} 
\label{Ham}
\end{equation}
where the indices $i$ and $j$ represent lattice sites, and the sum is
over all nearest neighbor pairs; $J$ is the coupling constant, $k_B$
is Boltzmann's constant, and $T$ is the temperature.
Thus, the partition function of the model is
\begin{equation}
Z_{\rm spin}=\int \prod_{<i,j>} \exp \left(\frac{J}{k_B T}  \,
\vec{s_{i}} \cdot \vec{s_{j}} \right) \prod_{k} d \vec{s_{k}}
\end{equation}
where the spins are normalized such that 
This model includes as special cases the Ising, the XY and the Heisenberg 
model, for $n=1$, $2$ and $3$, respectively. 

In the high-temperature limit, the bond weight
$w(\vec{s_{i}} \cdot \vec{s_{j}})$ reduces in first order to
$(1+x\vec{s_i}\cdot \vec{s_j})$  with $x=J/(k_BT)$. Thus the partition
function takes the form
\begin{equation}
Z_{\rm spin}=\int\prod_{\langle i, j\rangle }
(1+x\vec{s_i}\cdot \vec{s_j}) \prod_{k}d\vec{s_{k}}
\label{spinpartition}
\end{equation}
The bond weight still satisfies the O($n$) symmetry implied by 
Eq.~(\ref{Ham}). According to the assumption of universality, the
universality class of a phase transition is determined by only very
few parameters including the symmetry of the spin-spin interactions.
It is thus reasonable to expect \cite{N1} that the reduced Hamiltonian
that corresponds with Eq.~(\ref{spinpartition}), namely 
$H=-\sum_{<i,j>}\ln(1+x\vec{s_{i}}\cdot\vec{s_{j}})$
with $x=J/(k_BT)$ not necessarily small, still belongs to the
O($n$) universality class in two dimensions.

The O($n$) model (\ref{spinpartition}) on the honeycomb lattice
can be mapped on the O($n$) loop model \cite{DMNS} on the same lattice,
with a partition sum
\begin{equation}
Z_{\rm loop}=\sum_{G}x^{N_{b}}n^{N_{l}}
\label{graphpartition}
\end{equation}
where the graph $G$ covers $N_{b}$ bonds of the lattice,  and consists of 
$N_{l}$ closed, non-intersecting loops.

In the language of the $O(n)$ loop model, $x$ is the weight of a bond
visited by a loop, and $n$ is the loop weight, and no longer restricted
to be an integer.

The research of O($n$) models is a subject of a considerable history,
in which a prominent place is occupied by the exact results \cite{N1}
for the O($n$) loop model on the honeycomb lattice.
These results include the critical points for $-2\leq n\leq 2$, and the
temperature and the magnetic exponent. 

Also the geometric description of fluctuations at and near criticality 
has a long history, which goes back to the formulation of phase 
transitions in terms of the droplet model \cite{Fisher}.
For the $q$-state Potts model (for a review, see Ref.~\onlinecite{Wu}),
it was found that the fractal dimension of Kasteleyn-Fortuin (KF)
clusters \cite{KF} is equal to the magnetic scaling exponent $y_h$.
More generally, geometric Potts clusters can be defined by connecting 
neighboring, equal Potts spins by a bond percolation process. Several
new critical exponents were found by Coulomb gas and other 
methods \cite{CP, BKN, DBN,Con}. These exponents describe the geometric
properties and the renormalization flow of this model.

For the O($n$) loop model, the fractal dimension $d_l$ of the critical
loop gas has, to our knowledge, not yet been derived explicitly.
However, a clue can be obtained from the relation \cite{DBN,JS1} 
between the exponents describing random clusters of the tricritical
Potts model, and those describing Potts clusters in the critical Potts
model. From the equivalence of the critical O($n$)
model and the tricritical $q=n^2$-state Potts model \cite{N1}, one 
can thus associate the fractal dimension $d_l$ of the critical O($n$)
loop gas to the hull fractal dimension of critical Potts clusters.
The latter dimension was conjectured by Vanderzande \cite{V}. 

In the present paper, we focus on a fundamental non-thermodynamic scaling 
dimension behind the geometric properties of O($n$) loops which follows
from the known correspondence with the Coulomb gas \cite{N}, and relate it
to some exponents describing such properties. These exponents are exact
and include the fractal dimension $d_l$ of O($n$) loops at criticality. 
{}From a similar calculation, Duplantier and Saleur \cite{SD, DS} derived
the fractal dimension $d_a$ of the interior of these loops. 

While theoretical predictions are available, thus far there is no
numerical evidence in support of these, except for the Ising case $n=1$
or $q=2$ \cite{JS1}.  One of the reasons behind this situation may be
that the O($n$) partition sum contains the number of loops.
In the existing Monte Carlo algorithm for the loop model \cite{Wdyn},
the acceptance probability of a local update thus depends (for $n \neq 1$)
on the change of the number of loops due to the Monte Carlo move.
The determination of this number is a nonlocal task and requires a
number of operations that increases algebraically with the system size
$L$. Critical slowing down can make this situation even worse, so that
this way of simulation is restricted to rather small system sizes.

Until now, a sufficiently efficient Monte Carlo algorithm for
the O($n$) loop model has not been described.
Therefore, in this work, we develop a new Monte Carlo algorithm, which
is applicable to models with noninteger $n>1$, and uses local updates.
Although these updates typically lead to nonlocal modifications of the
loop connectivities, the number of operations required per update is
only of order one, and essentially independent of the system size. 

We then apply the algorithm to the critical O($n$) loop model and 
determine exponents of some geometric observables. 
The results agree with the theoretical predictions. 

The outline of the rest of this paper is as follows. In Sec. \ref{CG}
we show how a fundamental non-thermodynamic scaling dimension behind
some geometric properties of O($n$) loops can be derived exactly from
a mapping on the Coulomb gas, and how it relates to exponents describing
some geometric observables. Section \ref{MC} introduces the Monte Carlo
algorithm. In Sec. \ref{Sim} we apply the algorithm to the critical
O($n$) loop model and determine exponents of some geometric observables.

\section{Coulomb gas derivation and scaling formulas}
\label{CG}
It is well known that geometric and fractal properties of O($n$) loops
and various types of critical clusters can be analyzed by means of
a mapping on the Coulomb gas \cite{Kad,N}. A number of exact scaling
dimensions was obtained by this technique, see e.g.,
Refs.~\onlinecite{dN,DS,SD,BKN,N}. Here we base ourselves on these
analyses, which rely on a reformulation of correlation functions 
$g(r)$ in the model of interest in terms of the Coulomb gas.
The dimensions $X(e,m)$ associated with such correlation
functions are described by pairs of electric and magnetic charges,
$(e_0,e_r)$ and $(m_0,m_r)$, separated by a distance $r$:
\begin{equation}
X(e,m) = - \frac{e_0 e_r}{2 g} - \frac{m_0 m_r g}{2}
\label{Xem}
\end{equation}
where $g$ is the coupling constant of the Coulomb gas. For the critical
O($n$) model it is given by
\begin{equation}
g= 1+ \frac{1}{\pi} {\rm arccos} \frac{n}{2}
\label{g}
\end{equation}
where we use a normalization that is in agreement with earlier literature
but different from that used in
Ref.~\onlinecite{N} (our $g$ is four times smaller, and the charges
differ by a factor two such that the $X(e,m)$ are the same).

Let us now consider the correlation function at criticality describing
the probability that two lattice edges separated by a distance $r$ are
part of the same loop. It decays with an exponent $2X_l$ where $X_l$
is the O($n$) loop scaling dimension. The exponent $X_l$ is described
by a pair of magnetic charges $m_0=-m_r= 1$ and a pair of electric
charges $e_0=e_r=1-g$.  This leads to
\begin{equation}
X_l = 1-\frac{1}{2g}
\label{xl}
\end{equation}
This dimension $X_l$ is the renormalization exponent 
behind geometrical and fractal properties of O($n$) loops, just as the
renormalization exponents $X_t$ and $X_h$ determine the thermodynamic
singularities.

In another application of the Coulomb gas technique we can explore
corrections to scaling associated \cite{BKN} with the exponent
$X_{2l}$ that describes the decay of the probability that two O($n$) 
loops collide in two points separated by a distance $r$. The value of
this exponent is determined by electric charges as above and magnetic
charges $m_0=-m_r= 2$:
\begin{equation}
X_{2l} = 1-\frac{1}{2g}  + \frac{3g}{2}
\label{x2l}
\end{equation}
This exponent becomes marginal at $n=2$ and is thus expected to
lead to poor convergence of finite-size data near $n=2$.

The physical relevance of $X_l$ can be demonstrated by means of scaling
arguments. The probability $g_l(r)$ that two points at a distance $r$
lie on the same loop is, as given above, $g_l(r)\simeq a r^{-2X_l}$.
Let, at criticality, the fractal dimension of the loops be $d_l$.
Thus, under a rescaling by a linear factor $b$, the length $l$ of the
loop decreases by a factor $b^{d_l}$, and its density increases by a 
factor $b^{2-d_l}$. This determines the correlation in the rescaled
system as $g_l(r/b)\simeq a b^{4-2d_l} r^{-2X_l}$ which is, as specified
above, to be compared with $g_l(r/b)\simeq a b^{2X_l} r^{-2X_l}$.
It thus follows that the fractal dimension of loops is 
\begin{equation}
d_l = 2- X_{l}
\label{dl}
\end{equation}

Let $P_l(l)$ be the density of loops of length $l$. It is natural that,
at criticality, $P_l(l)$ depends algebraically on $l$, with an exponent
denoted as $p_l$: 
\begin{equation}
P_l(l) \propto l^{p_l} \, .
\label{ml}
\end{equation}
Under a rescaling by a linear factor $b$, the loop
density is affected for two reasons: first, the loops decrease in length
by a factor $b^{d_l}=b^{2- X_{l}}$; and second, the density increases
by a factor $b^2$ because the volume is reduced. Consistency requires
that $P_l(l)dl=b^{-2} P_l(l b^{2- X_{l}}) d(l b^{2- X_{l}})$ or
$P_l(l)=b^{- X_{l}} P_l(l b^{2- X_{l}})$.
The requirement $P_l(l) \propto l^{-p_l}$ yields
$l^{-p_l} = (l b^{-2+ X_{l}})^{-p_l} b^{-2+ X_{l}-2}$. Matching the exponents
shows that
\begin{equation}
p_l = 1+\frac{2}{2- X_{l}}
\label{pl}
\end{equation}
Since scaling implies that the divergence of the expectation value of the
linear size of the largest loop goes as $(x_c-x)^{-\nu}=(x_c-x)^{-1/y_t}$
when the critical point is approached, and the actual length $l_{\rm max}(x)$ 
as expressed in lattice edges behaves as a power $2-X_l$ of the linear
size, it follows that the largest loop length diverges as
\begin{equation}
l_{\rm max}(x) \propto  (x_c-x)^{(X_l-2)/y_t}
\label{lm}
\end{equation}

For $L \to \infty$ and $x>x_c$, there exists an infinite spanning loop.
Under a rescaling by
a linear scale factor $b$ its density increases by a factor $b^{X_l}$
while, as usual, the temperature field $t\propto x-x_c$ scales as
$t \to t' = b^{y_t} t$. The fraction $s_l(x-x_c)$ of the edges covered
by the spanning loop scales as $s_l(b^{y_t}(x_c-x))=b^{X_l} s_l((x_c-x)$.
The choice $b=(x_c-x)^{-1/y_t}$ leads to a constant on the left hand
side of this equation, and after substitution in the right hand side,
the scaling behavior follows as
\begin{equation}
s_l(x_c-x)\propto (x_c-x)^{X_l/y_t} 
\label{lm1}
\end{equation}
The finite-size dependence of the similar fraction $s_L$ of a system 
with finite size $L$ at criticality can simply be found by rescaling
the system to a given size, say 1. This leads to
\begin{equation}
s_L  \propto L^{-X_l}
\label{ss}
\end{equation} 
Including a correction to scaling, we may modify this into
\begin{equation}
s_L =a L^{-X_l}(1+b L ^{y_i}+\cdots)
\label{ssc}
\end{equation}
where $y_i=2-X_{2l}$ is a candidate for the leading correction
exponent \cite{BKN}, and $a$ and $b$ are unknown amplitudes.

In analogy with magnetic systems, a susceptibility-like quantity $\chi_l$
can be defined on the basis of the distribution of the loop sizes as
\begin{equation}
\chi_l\equiv \sum_{l=1}^{l_{\rm max}} P_l(l) l^2
\label{chid}
\end{equation}
According to the aforementioned scaling behavior, the largest loop
in a critical system of finite size $L$ has a length scale
$l_{\rm max}\propto L^{2-X_l}$. Thus
\begin{equation}
\chi_l(L) \propto \sum_{l=1}^{L^{2-X_l}} l^{2-p_l} 
\label{chil}
\end{equation}
Substitution of $p_l = 1+ 2 /(2- X_{l})$ yields
\begin{equation}
\chi_l(L) \propto L^{2-2X_l}
\label{chis}
\end{equation}
Again, we can include corrections to scaling 
\begin{equation}
\chi_l(L) = c L^{2-2X_l}(1 + d L^{y_i} +\cdots)
\label{chisc}
\end{equation}
with $y_i$ as mentioned above, and $c$ and $d$ are unknown constants. 

Another correlation function of interest describes the probability that
two sites on the dual triangular lattice separated by a distance $r$
sit inside the same loop (i.e., not separated by any loop of the model).
The exponent $X_a$ describing the decay of this function at criticality
was derived by Duplantier and Saleur \cite{DS} as $X_a=1-g/2-3/(8g)$.
The fractal dimension of the interior of O($n$) loops is therefore 
$d_a=2-X_a=1+g/2+3/(8g)$. The area inside a loop  does not include
the area inside loops enclosed by that loop.  The exponent $X_a$ thus
also determines the finite-size scaling of the spanning loop. We are 
therefore interested in the fraction $s_a$ of the dual lattice sites 
that sit inside the spanning loop.
Scaling indicates that this quantity is subject to the following
finite-size behavior:
\begin{equation}
s_a(L) \propto L^{-X_a}
\label{sis}
\end{equation}
We furthermore define
another susceptibility-like quantity $\chi_a$ on the basis of the 
distribution $P_a(a)$ of the area $a$ (expressed in the number of
enclosed sites of the dual lattice) of the interior of the loops as
\begin{equation}
\chi_a(L) \equiv  L^{-2} \sum_a P_a(a) a^2
\label{xis}
\end{equation}
 which is expected to scale as
\begin{equation}
\chi_a(L) \propto L^{2-2X_a}
\label{chiis}
\end{equation}

\section{Monte Carlo Algorithm}
\label{MC}
In the existing Monte Carlo algorithm for the loop model \cite{Wdyn},
the local updates require time-consuming nonlocal operations as 
explained above. To get rid of these we adopt the following procedure.

\begin{figure}
\centering
\includegraphics[scale=0.6]{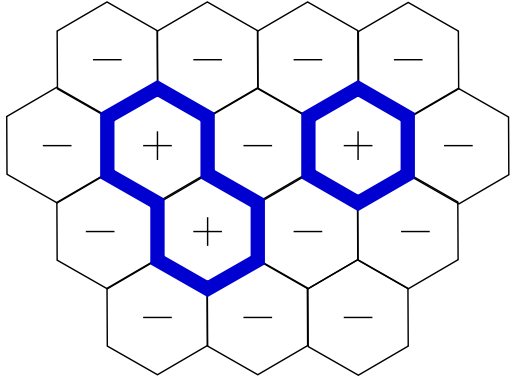}
\caption{Representation of a loop configuration with Ising spins}
\label{cfg1}
\end{figure}

As a first step of such an algorithm for the simulation of the O($n$)
loop model on the
honeycomb lattice, we represent the loop configuration by means of 
Ising spins on the dual lattice, which is triangular. The loops are just
the interfaces between neighboring spins of a different sign.
We restrict ourselves to systems with periodic boundary conditions, so
that the interfaces indeed form a system of closed, nonintersecting loops 
loops on the honeycomb lattice.
This is illustrated in Fig.~\ref{cfg1} using a loop configuration which
consists of 2 loops and 16 bonds, shown as bold lines. This graph
contributes a weight $x^{16}n^{2}$ to the partition function. The loops
can of course be represented by two opposite Ising spin configuration,
but this degeneracy has no further consequences for our line of argument.

We now show how one can update the loop configuration by means of local
Metropolis-type updates of the Ising spins representing the loop
configuration. It works only for $n \geq 1$.
One step of importance sampling is realized by the following operations,
which are to be repeated cyclically:
\begin{enumerate}
\item
For each loop, assign its color to be either 'green' with probability
$1/n$ or 'red' with the remaining probability $1-1/n$.
\item
Randomly select an Ising spin on the dual lattice. 
\item
Check if the spin is adjacent to a red loop segment. If so, do
nothing; if not, update the spin using the Metropolis probabilities,
with Ising couplings $K$ determined as $e^{-2K}=x$.
\item
Repeat steps 2 and 3 until the number of update attempts is equal to
the number of sites of the dual lattice.
\item
Perform a sweep through the Ising system to find all loops.
\item
Repeat steps 1 to 5 a fixed number $n_s-1$ times.
\item
Sample the data of interest from the loop configuration.
\end{enumerate}
A Monte Carlo run consists of many of these cycles, each of which thus
includes $n_s$ Metropolis sweeps, new random assignments of loop colors,
and the data sampling procedure.
Each of these steps satisfies the conditions of detailed balance,
so that the algorithm should indeed generate configurations in accordance 
with Boltzmann statistics. Tests confirm that the simulation results
agree with those of the existing algorithm \cite{Wdyn}.
Since this algorithm assigns colors to the loops, we refer to it
as "coloring algorithm".

\section{Simulation}
\label{Sim}
With the help of the algorithm described above, we simulated the O($n$)
model on the honeycomb lattice at the exactly known \cite{N1} critical
points $x_c=(2+(2-n)^{1/2})^{-1/2}$ in order to check the theoretical
predictions described in Sec.~\ref{CG}.

We first applied our algorithm to the case $n=1.5$, using $12$ system
sizes in the range $8 \leq L \leq 128$,  where $L$ is the linear size
of the dual triangular lattice, and periodic boundary conditions.

For each system size, a run was executed with a length of $4\times 10^7$
Monte Carlo cycles after equilibration of the system. Each cycle 
included $n_s=5$ sweeps and loop formation steps, and one sampling as 
described above.  
Statistical errors were estimated by means of binning in 2000 partial
results.

\begin{figure}
\centering
\includegraphics{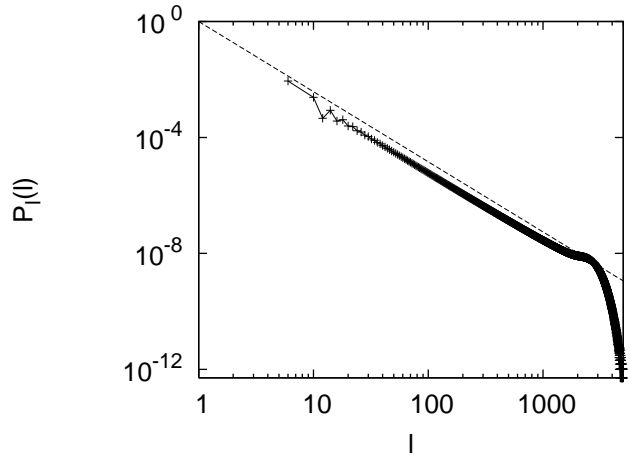}
\caption{Distribution $P_l$ of loops of length $l$ on logarithmic
scales, for the critical O($n$) model with $n=1.5$ and size $L=128$.
The dashed line shows a power law decay with exponent -2.42198318,
which is the theoretical asymptotic value for the infinite system. }  
\label{ml15}
\end{figure}

The data sampling included the density $P_l(l)$ of loops of length $l$.
The results for the system of size $L=128$ are shown in Fig.~\ref{ml15}. 
It displays a substantial interval of algebraic decay, as described by
Eq.~(\ref{ml}).

The fractions $s_l$ and $s_a$, and the susceptibility-like quantities
$\chi_l$ and $\chi_a$ were also sampled. The finite-size dependence
of these quantities at criticality is shown in Table \ref{tablen15}.

\begin{table}[table1]
\caption{numerical data for $s_l(L), \chi_{l}(L)$ and $s_a(L),\chi_a(L)$ for 
different system sizes $L$ at the critical point of  O($n=1.5$) model}
\label{tablen15}
\begin{tabular}{l|lr|lr|lr|lr}
\hline
$L $     & $s_l$    &    &$\chi_{l}$&   & $s_a$  &     &$\chi_a$&     \\
\hline
    8    &  0.12907 &(2) &  0.8648&(1)  & 0.81028&(3)  &   2.166&(1)  \\
   16    &  0.08611 &(2) &  2.1984&(3)  & 0.76386&(5)  &   8.439&(6)  \\
   24    &  0.06785 &(3) &  3.4141&(6)  & 0.73829&(8)  &  18.39 &(2)  \\
   32    &  0.05718 &(3) &  4.5578&(8)  & 0.72108&(9)  &  31.66 &(5)  \\
   40    &  0.05006 &(4) &  5.6466&(11) & 0.70812&(11) &  48.15 &(9)  \\
   48    &  0.04504 &(4) &  6.6993&(14) & 0.69729&(13) &  68.11 &(14) \\
   56    &  0.04107 &(4) &  7.7199&(23) & 0.68869&(16) &  90.80 &(24) \\
   64    &  0.03793 &(4) &  8.7142&(25) & 0.68130&(15) & 116.3  &(3)  \\
   80    &  0.03324 &(4) & 10.635 & (3) & 0.66899&(19) & 176.6  &(6)  \\
   96    &  0.02987 &(4) & 12.477 & (6) & 0.65895&(22) & 249.0  &(10) \\
  112    &  0.02723 &(5) & 14.273 & (6) & 0.65095&(26) & 330.7  &(14) \\
  128    &  0.02513 &(5) & 16.027 & (8) & 0.6441 & (3) & 422.4  &(23) \\
\hline
\end{tabular}
\end{table}

\begin{figure}
\centering
\includegraphics{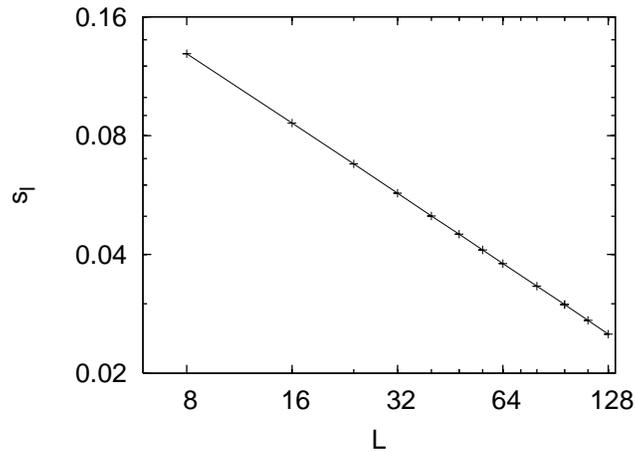}
\caption{Fraction $s_l$ of lattice edges covered by the spanning loop,
versus system size $L$ for the critical O($n$) model with $n=1.5$,
on logarithmic scales. The curve is added as a guide to the eye, and 
estimated error bars are smaller than the size of the symbols.}
\label{s1.5}
\end{figure}

\begin{figure}
\centering
\includegraphics{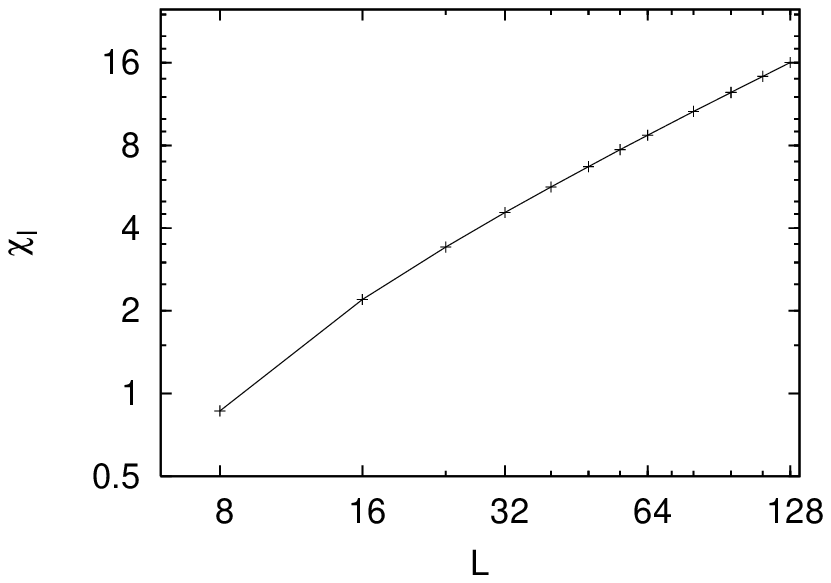}
\caption{Susceptibility-like quantity $\chi_l$ versus system size $L$
for the critical O($n$) model with $n=1.5$ on logarithmic scales. The
curve is added as a guide to the eye, and estimated error bars
are smaller than the size of the symbols.}

\label{chi1.5}
\end{figure}

\begin{figure}
\centering
\includegraphics{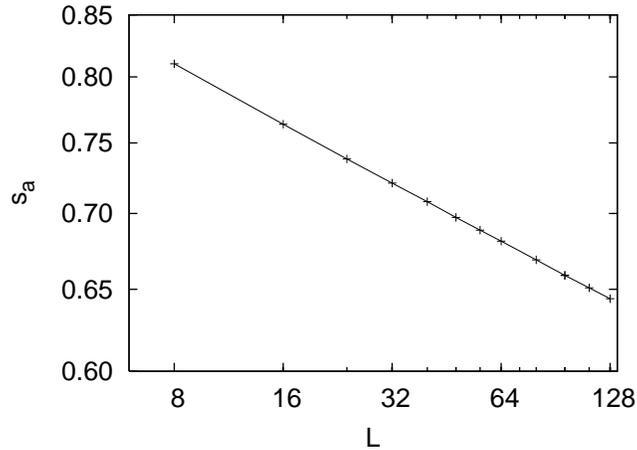}
\caption{Fraction $s_a$  of the number of dual lattice sites
inside the spanning loop, versus system size $L$ for the critical
O($n$) model with $n=1.5$, on logarithmic scales. The curve is added
as a guide to the eye, and estimated error bars are smaller than the
size of the symbols.}
\label{si1.5}
\end{figure}

\begin{figure}
\centering
\includegraphics{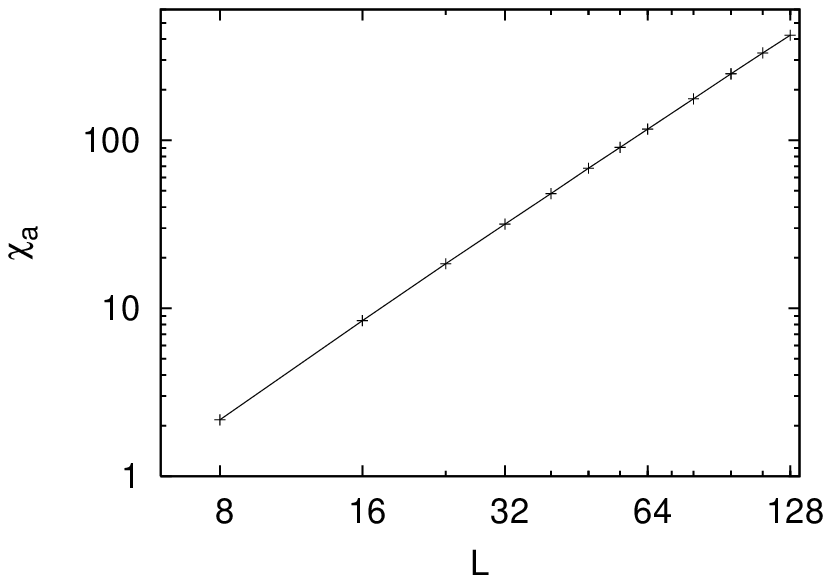}
\caption{Susceptibility-like quantity $\chi_a$ versus system size $L$
for the critical O($n$) model with $n=1.5$ on logarithmic scales. The
curve is added as a guide to the eye, and estimated error bars
are smaller than the size of the symbols.}
\label{chii1.5}
\end{figure}

These quantities are well described by power laws as a function of the
lattice size $L$  for sufficiently large $L$. This is just as expected 
on the basis of finite-size scaling as expressed by Eqs.~(\ref{ss}),
(\ref{chis}), (\ref{sis}) and (\ref{chiis}). This behavior is illustrated
in Figs.~\ref{s1.5}, \ref{chi1.5}, \ref{si1.5} and \ref{chii1.5}.

Using the nonlinear Levenberg-Marquardt least-squares algorithm, we 
fitted for the exponents and amplitudes
according to the finite-size-scaling formulas including correction terms,
as given in Eqs.~(\ref{ssc}), (\ref{chisc}), (\ref{sis}) and (\ref{chiis}). 
Thus we can numerically determine $X_l$, $X_a$, and thereby the
fractal dimension $d_l$ of the spanning loop, and the 
fractal dimension $d_a$ of interior of the spanning loop.
Comparing the residual $\chi^2$ of the fits with the number of degrees
of freedom, we find satisfactory fits including all system sizes $L\geq 8$.
We obtain $X_l=0.593 \,(2)$ from the fit of $s_l(L)$ 
and $X_l=0.595 \,(2)$ from the fit of $\chi(L)$.
These  results are consistent with the theoretical value
$X_l \approx 0.593513601$.
The fit of $s_a(L)$ yields $X_a=0.080 \,(1)$. From the fit of $\chi_a(L)$, 
we obtain $2-2 X_a=1.84\,(1)$, or $X_a=0.080\,(5)$. These results agree 
well with the theoretical prediction $X_a \approx 0.0801085234$.

In these fits, the correction-to-scaling exponent $y_i$ was left free.
The fits suggest that the exponent of the dominant correction to
scaling does not assume the expected value $2-X_{2l}=-0.43859$, but
instead $y_i=-0.75 \pm 0.10$.
On the other hand, we have also performed similar simulations of the O($1$) 
model at the critical point. The fractions $s_l$ and $s_a$, and the
susceptibility-like quantities $\chi_l, \chi_a$ were sampled for
several system sizes. The results are shown in Tables \ref{tablen1}. 
Least-squares fit results agree well with the theoretical prediction, 
as listed in Table \ref{fitresult}. 
For the O(1) model, only the data for $\chi_l$ allowed a reasonably accurate
estimate of the leading correction-to-scaling exponent. In this case,
we find $y_i=-0.628\,(7)$, in a good agreement with the expected value
$2-X_{2l}=-0.625$. We thus have fixed $y_i$ at this value in the other fits.

Furthermore we performed similar simulations of the O($n$) models with
$n=\sqrt{2}$, $n=\sqrt{3}$ and $n=2$ at their critical points as given in
Ref.~\onlinecite{N1}.  The  fractions $s_l$ and $s_a$, and the
susceptibility-like quantities $\chi_l, \chi_a$ were sampled for
several system sizes. The results are shown in Tables \ref{tablesqrt2},
\ref{tablensqrt3} and \ref{tablen2}.
Also these quantities appear to depend algebraically on the system size
$L$, in agreement with the finite-size scaling equations
(\ref{ss}), (\ref{chis}), (\ref{sis}) and (\ref{chiis}). 

\begin{table}[htbp]
\caption{Numerical data for $s_l(L)$, $\chi_{l}(L)$, $s_a(L)$ and
$\chi_a(L)$ for several system sizes $L$ at the critical point of the
O($n=1$) model.}
\label{tablen1}
\begin{center}
\begin{tabular} {l|lr|lr|lr|lr}
\hline
    $L $ & $s_l$   &    &$\chi_{l}$&   &    $s_a$  &    & $\chi_a$&       \\
\hline
    8    &  0.09126&(2) &  0.43532&(8) &   0.88355 &(2) &   1.1563& (6)   \\
   16    &  0.06015&(2) &  1.1859 &(2) &   0.85090 &(5) &   4.580 & (5)   \\
   24    &  0.04678&(2) &  1.8531 &(3) &   0.83257 &(7) &  10.108 &(16)   \\
   32    &  0.03911&(2) &  2.4658 &(4) &   0.81987 &(9) &  17.682 &(39)   \\
   40    &  0.03405&(3) &  3.0413 &(6) &   0.81015 &(12)&  27.27  & (8)   \\
   48    &  0.03040&(3) &  3.5901 &(8) &   0.80234 &(14)&  38.73  &(13)   \\
   56    &  0.02754&(3) &  4.1149 &(12)&   0.79617 &(15)&  51.68  &(19)   \\
   64    &  0.02539&(4) &  4.6235 &(14)&   0.79038 &(20)&  66.98  &(32)   \\
   80    &  0.02214&(5) &  5.5903 &(24)&   0.78077 &(27)& 104.0   & (7)   \\
   96    &  0.01968&(5) &  6.5121 &(33)&   0.77384 &(28)& 145.3   &(11)   \\
  112    &  0.01781&(6) &  7.4027 &(42)&   0.76799 &(35)& 193.0   &(18)   \\
  128    &  0.01645&(6) &  8.2442&(43) &   0.76213 &(42)& 253.2&(26)    \\
\hline
\end{tabular}
\end{center}
\end{table}

\begin{table}[htbp]
\caption{Numerical data for $s_l(L)$, $\chi_{l}(L)$, $s_a(L)$ and 
$\chi_a(L)$ for several system sizes $L$ at the critical point of the
O($n=\sqrt{2}$) model.}
\label{tablesqrt2}
\begin{center}
\begin{tabular} {l|lr|lr|lr|lr}
\hline
    $L $ & $s_l$    &   & $\chi_{l}$&  & $s_a$      &   & $\chi_a$& \\
\hline
    8    &  0.12186 &(2) &   0.7813 &(1) &  0.82429 &(3)&    1.953&(1)    \\
   16    &  0.08104 &(2) &   1.9989 &(3) &  0.78004 &(5)&    7.651&(7)    \\
   24    &  0.06361 &(3) &   3.1027 &(5) &  0.75586 &(8)&   16.65 &(2)    \\
   32    &  0.05358 &(3) &   4.1362 &(7) &  0.7392  &(1)&   28.81 &(5)    \\
   40    &  0.04693 &(4) &   5.1183 &(11)&  0.7264  &(1)&   44.12 &(9)\\
   48    &  0.04206 &(4) &   6.0630 &(16)&  0.7164  &(2)&   62.14 &(18)    \\
   56    &  0.03835 &(5) &   6.9786 &(18)&  0.7081  &(2)&   82.86 &(22)    \\
   64    &  0.03536 &(4) &   7.8623 &(22)&  0.7011  &(2)&  106.5  &(3)    \\
   80    &  0.03096 &(5) &   9.580  & (4)&  0.6890  &(2)&  162.4  &(6)    \\
   96    &  0.02769 &(5) &  11.222  & (5)&  0.6800  &(3)&  227.0  &(12)    \\
  112    &  0.02532 &(6) &  12.825  & (7)&  0.6716  &(3)&  304.4  &(18)    \\
  128    &  0.02338 &(6) &  14.370  & (9)&  0.6649  &(4)&  390.7  &(25)    \\
\hline
\end{tabular}
\end{center}
\end{table}

\begin{table}[htbp]
\caption{Numerical data for $s_l(L)$, $\chi_{l}(L)$, $s_a(L)$ and
$\chi_a(L)$ of the critical O($n=\sqrt{3}$) model for several system
sizes $L$.}
\label{tablensqrt3}
\begin{center}
\begin{tabular} {l|lr|lr|lr|lr}
\hline
    $L $ & $s_l$    &    &$\chi_{l}$&     & $s_a$    &    & $\chi_a$ &      \\
\hline
    8    &  0.15197 &(2) &   1.1247 &(2)  &  0.76686 &(3) &    2.891 &(1)   \\
   16    &  0.10286 &(2) &   2.8312 &(3)  &  0.71434 &(5) &   11.150 &(6)   \\
   24    &  0.08169 &(3) &   4.4245 &(7)  &  0.68606 &(6) &   24.00  &(2)   \\
   32    &  0.06929 &(4) &   5.9453 &(12) &  0.66677 &(9) &   41.16  &(5)   \\
   40    &  0.06111 &(4) &   7.4130 &(17) &  0.6522  &(1) &   62.44  &(9)   \\
   48    &  0.05500 &(4) &   8.8463 &(23) &  0.6408  &(1) &   87.2 1 &(13)  \\
   56    &  0.05035 &(4) &  10.241  & (3) &  0.6314  &(1) &  115.7   &(2)   \\
   64    &  0.04665 &(4) &  11.611  & (4) &  0.6233  &(1) &  148.0   &(3)   \\
   80    &  0.04112 &(5) &  14.272  & (6) &  0.6098  &(2) &  223.7   &(6)   \\
   96    &  0.03701 &(4) &  16.885  & (7) &  0.5993  &(2) &  311.0   &(9)   \\
  112    &  0.03387 &(4) &  19.407  & (9) &  0.5906  &(2) &  411.4   &(12)  \\
  128    &  0.03153 &(5) &  21.865  &(11) &  0.5824  &(2) &  531.2   &(19)  \\
\hline
\end{tabular}
\end{center}
\end{table}

\begin{table}[htbp]
\caption{Numerical data for $s_l(L)$, $\chi_{l}(L)$, $s_a(L)$ and
$\chi_a(L)$ of the critical O($n=2$) model for several system
sizes $L$.}
\label{tablen2}
\begin{center}
\begin{tabular} {l|lr|lr|lr|lr}
\hline
    $L $ & $s_l$  &    & $\chi_{l}$&  & $s_a$   &    &$\chi_a$&       \\
\hline
    8  & 0.21534  &(3) & 1.7038 &(3)  & 0.66465 &(3) &  5.082 &(1)    \\
   16  & 0.15273  &(2) & 4.4449 &(7)  & 0.60441 &(3) & 18.981 &(6)    \\
   24  & 0.12479  &(2) & 7.2121 &(11) & 0.57283 &(4) & 39.965 &(16)   \\
   32  & 0.10812  &(2) & 9.9856 &(17) & 0.55170 &(4) & 67.330 &(29)   \\
   40  & 0.09674  &(2) &12.762  &(2)  & 0.53600 &(5) &101.00  &(5)    \\
   48  & 0.08832  &(2) &15.544  &(3)  & 0.52358 &(5) &139.39  &(7)    \\
   56  & 0.08177  &(3) &18.328  &(4)  & 0.51336 &(6) &183.47  &(11)    \\
   64  & 0.07650  &(2) &21.106  &(4)  & 0.50468 &(5) &232.70  &(14)    \\
   80  & 0.06844  &(3) &26.676  &(7)  & 0.49057 &(7) &345.66  &(30)    \\
   96  & 0.06243  &(2) &32.251  &(9)  & 0.47944 &(6) &476.81  &(37)    \\
  112  & 0.05783  &(3) &37.828  &(11) & 0.47013 &(7) &626.4   &(5)    \\
  128  & 0.05410  &(3) &43.392  &(13) & 0.46223 &(7) &793.1   &(7)    \\
  192  & 0.04418  &(3) &65.680  &(26) & 0.43916 &(10)&1620    &(2)    \\

\hline
\end{tabular}
\end{center}
\end{table}

Using finite size scaling and the Levenberg-Marquardt algorithm, we 
determined the exponents $X_l$ and $X_a$. The results
are summarized in Table \ref{fitresult}.

\begin{table}[htbp]
\caption{Results for the exponents $X_l$ and $X_a$ for five O($n$)
models, as obtained from least-squares fits as described in the text.}
\label{fitresult}
\begin{center}
\begin{tabular} {l|lr|l|lr|lr|l|lr}
\hline
$n$&$X_l$       &  &$X_l$   &$2-2X_l$& &$X_a$    &   &$X_a$   & $2-2X_a$ &\\
&from $s_l$  &  &theory&from $\chi_l$& &from $s_a$&&theory &from $\chi_a$&\\
\hline
$1$      &0.625 &(1) & 5/8   & 0.73  &(2)&0.0518 &(3)&0.05208 &1.89 &(1)\\
$\sqrt2$ &0.599 &(1) & 3/5   & 0.796 &(5)&0.075  &(1)& 3/40   &1.86 &(1)\\
$1.5   $ &0.595 &(3) &0.5935 & 0.810 &(4)&0.080  &(1)&0.08011 &1.84 &(1)\\
$\sqrt3$ &0.571 &(1) &0.5714 & 0.856 &(1)&0.095  &(1)&0.0952  &1.81 &(1)\\
$2$      &0.4997&(3) & 1/2   & 0.996 &(5)&0.1249 &(4)&1/8     &1.747&(2)\\ 
\hline
\end{tabular}
\end{center}
\end{table}
\section{Discussion}
\label{Dis}
The time-consuming character of simulations of loop models is due to the
nonlocal character of the loops. We have reduced this problem by splitting 
the loop weight $n$ in two parts, namely $n-1$ and $1$. Proper summation
over both contributions is done by assigning color variables to the
loops; a sum on all color variables is included in the partition sum.
The algorithm treats these color variables as dynamical variables which
are updated by the Monte Carlo process.
The idea to use an additional color variable for each loop
was already used in a context unrelated to Monte Carlo methods,
e.g. in Ref.~\onlinecite{sq}.  The presence of loops of weight 1
in such a configuration then leads, at least locally, to a system
that behaves precisely as an Ising configuration. Thus, the system may
be locally updated by means of Metropolis-like Monte Carlo steps.
Care should be taken to leave the loops of weight $n-1$ unchanged,
because it would violate the condition of detailed balance.
The procedure given in Sec.~\ref{MC} satisfies this constraint.

The development of this coloring algorithm was motivated by the
possibility to further explore the physics of O($n$) models.
In the course of this work, we realized that it should be possible to
construct an even more efficient algorithm of a `cluster' nature.
Algorithms of this type will be presented elsewhere \cite{DGBca}.
The `coloring' trick is only useful for $n>1$. For $0<n<1$ the
existing algorithm \cite{Wdyn} is, although relatively inefficient,
still applicable.

For the interpretation of the simulation results, it is relevant that
we are using periodic boundary conditions, and that the mapping between
loop and Ising configurations imposes a condition of `evenness' on the 
loop configurations: a path spanning the periodic boundaries must 
have an even number of intersections with a loop. Therefore, the
statistical ensemble generated by the algorithm does not coincide
with that of Eq.~(\ref{graphpartition}). The difference is related
to the boundary conditions and is expected to modify the finite-size
behavior, but should vanish in the thermodynamic limit.

The present work is restricted to the `even' loop configurations.
It is, however, possible to simulate `odd' loop configurations by 
introducing a `seam' on the dual lattice, a vertical column of
horizontal antiferromagnetic Ising bonds spanning the system.
For these antiferromagnetic bonds we use the rule that there is a loop
segment if and only if the two associated dual spins are of the same
sign. Horizontal and vertical seams can be introduced independently, as
prescribed by the class of loop configurations that is to be sampled.

\acknowledgments
We wish to thank Prof. B. Nienhuis for many valuable discussions.
This work was supported by by the National Science Foundation of China
under Grant \#10105001, and by the Lorentz Fund (Leiden University).

\newpage

\end{document}